\documentclass{iopart}
\usepackage{graphicx}
\usepackage{epsf,amsfonts,amssymb,amsbsy}
\begin{document}
\title[Vector field and rotational curves]
{Vector field and rotational curves in dark galactic halos}
\author{V.V.Kiselev*\dag}
\address{*
\ Russian State Research Center ``Institute for
High Energy
Physics'', 
Pobeda 1, Protvino, Moscow Region, 142281, Russia}
\address{\dag\ 
Moscow
Institute of Physics and Technology, Institutskii per. 9,
Dolgoprudnyi Moscow Region, 141700, Russia}
\ead{kiselev@th1.ihep.su}
\begin{abstract}
We study equations of a non-gauge vector field in a spherically
symmetric static metric. The constant vector field with a scale
arrangement of components: the temporal component about the Planck
mass $m_{\rm Pl}$ and the radial component about $M$ suppressed
with respect to the Planck mass, serves as a source of metric
reproducing flat rotation curves in dark halos of spiral galaxies,
so that the velocity of rotation $v_0$ is determined by the
hierarchy of scales: $ \sqrt{2}\, v_0^2= M/m_{\rm Pl}$, and $M\sim
10^{12}$ GeV. A natural estimate of Milgrom's acceleration about
the Hubble rate is obtained.
%
\end{abstract}


\section{Introduction}
In contrast to a scalar field appropriately elaborated in the
cosmology lately, a vector field has remained beyond intensive
studies of astro-particle problems.

Indeed, a slowly rolling scalar field drives an inflation
\cite{lectures}, which produces a correct spectrum of
inhomogeneous perturbations in a matter density as was observed in
an anisotropy of cosmic microwave background \cite{WMAP}. Next, a
scalar field naturally gives a state with a negative pressure, a
quintessence \cite{Q,state,trod}, so that it can cause an
accelerated expansion of Universe \cite{SNe} with a dynamical dark
energy. Moreover, some authors try to ascribe a dark matter in
halos of galaxies \cite{lect,univers,rev} to a scalar field, too,
though properties of such the fields at the galaxy scales should
be rather different from those of quintessence at the cosmic scale
\cite{expl,qdm,me,fay}. So, the scalar fields provide a
preferable, more or less successful theoretical treatment of
important astro-particle phenomena\footnote{An example of
alternative description due to a modification of gravity is given
in \cite{Moffat}.}. Though some tuning procedures might be
required due to a arbitrariness of scalar potentials. So, the
scalar particle of quintessence should have a mass of Hubble
constant scale, for instance. In addition, the dark matter
distribution in the spiral galaxies falls as $1/r^2$, which could
be a problem for a theoretical explanation in terms of both a
scalar field or exotic weak interacting particles.

However, as was recently observed \cite{vecquin}, a vector field
dynamics possesses features, which make it attractive for the
cosmological studies. So, in the evolution of flat isotropic
homogeneous universe a vector field gains a dynamical mass
proportional to the Hubble constant at any, even trivial,
potential\footnote{Note, that a reasonable form of potential for a
vector field is more strictly constrained than that of a scalar
field.}. Therefore, we do not need any synthetic assumptions to
get such the characteristic property in the sector of scalar
fields\footnote{The interacting scalar fields could give a
dynamical mass of Hubble scale as was shown in \cite{Lyth}.}, but
instead it is enough to introduce an isotropic vector field in the
cosmology, and covariant derivatives automatically generate such
the dynamical mass term.

We study a simple Lorentz-invariant form of lagrangian for a
vector field interacting gravitationally only\footnote{A gauge
vector field with a global symmetry was investigated in
\cite{bertl}, while a non-linear electrodynamics, which leads to
an acceleration of universe, was studied in \cite{berg}.},
\begin{equation}\label{v-lagr}
    {\cal L}_\textsl{v} =\xi\,\frac{1}{2}\,g^{\mu\nu}(\nabla_\mu\phi^m)\,
    (\nabla_\nu\phi^n)\,g_{mn}-V(\phi^2),
\end{equation}
where $\xi$ is a vector field signature\footnote{The signature of
metric is assigned to $(+,-,-,-)$.}, that can be normal ($\xi=-1$)
or phantomic ($\xi=+1$), respectively\footnote{Note that the
four-vector is generically composed by spin-1 and spin-0
components (see ref. \cite{ahu} for discussion on a covariant
object decomposition into components with definite spins).}.

Since a temporal component of the four-vector has a negative sign
of its kinetic term at the normal signature $\xi=-1$, we will not
treat (\ref{v-lagr}) as a fundamental lagrangian in a sense of
axiomatic field theory, but we address it as an effective
phenomenological lagrangian with a phantom: the field possessing
the negative kinetic term. So, we accept the following
\textbf{postulate of validity} for the phantom phenomenology:
\begin{itemize}
    \item the phantom component is not observable in the flat space-time,
    i.e. it does not interact with matter field (one can use a standard technique
    in order to introduce a gauge invariant lagrangian for a vector field);
    \item in gravity, the phantom is valid in regions \textit{beyond
    physical singularities}, if exist.
\end{itemize}
In this respect, a motivation for the lagrangian of (\ref{v-lagr})
could be manyfold.

First, a phenomenology of a scalar quintessence shows that a state
parameter of the quintessence, i.e. the ratio of a pressure to a
density of energy $w=p/\rho$, can take values less than $-1$
\cite{state}:
\begin{equation*}
    w<-1,
\end{equation*}
that, probably, puts the quintessence into the phantom stage: the
field with a negative sign of kinetic energy
\cite{trod,BigRip,johri}. The lagrangian of vector field in
(\ref{v-lagr}) {\it naturally} contains a phantom component, the
temporal component at normal signature $\xi=-1$. An important
physical difference from the case of scalar phantom is a
non-trivial covariant derivative of the vector field.

However, the negative kinetic energy of a fundamental field is not
restricted from below, so that the phantom cannot be considered as
a fundamental field. Such the treatment of scalar phantom involves
a cut-off for the kinetic term or equivalently higher derivative
terms providing a low boundary of negative energy. For example, we
can add a term of the form
\begin{equation*}
    \delta {\cal L}=\frac{1}{\Lambda^4}\,\dot \phi^4,\qquad
    \dot\phi =\frac{{\rm d}\phi}{{\rm d} t},
\end{equation*}
which guarantees a `stabilization' of negative kinetic term ${\cal
L}_0=-1/2\,\dot \phi^2$ for the scalar phantom. Nevertheless,
greater characteristic time-scale of field changes with respect to
the inverse stabilization scale $1/\Lambda$ is more accurate
leading approximation by the small negative kinetic term. So, we
can expect that at cosmological and galactic scales the effective
phantom theory with negative kinetic term could be rather sound,
if the stabilization scale is microscopic enough. Decay
instabilities of phantom in the presence of several phantoms were
considered in \cite{trod}. Thus, standard arguments justifying the
study of scalar phantoms are applicable for the introduction of
lagrangian (\ref{v-lagr}) for the vector field.

Second, theories with extra dimensions generate cosmological
equations including specific terms quadratic in energy-momentum
tensors of matter propagating in 4 di\-men\-sions in addition to
ordinary energy-momentum term \cite{Brax}. This fact can be
con\-si\-dered as an effective redefinition of energy-momentum
tensor:
\begin{equation*}
    \rho\rightarrow \rho+\frac{1}{2\sigma}\,\rho^2,
\end{equation*}
where $\sigma$ is a bare cosmological constant (a brane tension).
So, the negative kinetic term of phantom should generate the term
quadratic in the kinetic energy, which effectively reproduces the
cut-off mentioned above.

Third, let us show that in contrast to a naive expectation a free
phantom interacting with a gravity only, cannot propagate in a
homogeneous isotropic space-time at all. Indeed, in the
Friedmann--Robertson--Walker metric
\begin{equation}\label{FRW}
    {\rm d}s^2 ={\rm d}t^2-a^2(t)\,[{\rm d}r^2+r^2{\rm
    d}\theta^2+r^2\sin^2\theta\,{\rm d}\varphi^2],
\end{equation}
we get the following independent field equations for the
time-dependent homogeneous scalar phantom $\phi$:
\begin{equation}\label{no-go}
    H^2=\frac{8\pi
    G}{3}\,\left(-\frac{1}{2}\,\dot\phi^2+V_0\right),\qquad \ddot
    \phi+3H\,\dot\phi=0,\qquad H\equiv \frac{\dot a}{a},
\end{equation}
where we have added a cosmological constant $V_0$ for reference.
At flat limit of $V_0=0$ we get evident condition
\begin{equation*}
    H^2\geqslant 0\quad\Rightarrow\quad \dot \phi\equiv 0,
\end{equation*}
and the phantom is a constant field, which does not evolve (or
propagate). Moreover, in the case of positive cosmological
constant $V_0>0$, the `pathological' kinetic energy is
automatically restricted
\begin{equation*}
    \dot \phi^2 \leqslant 2V_0.
\end{equation*}
Such the kind of the restriction remains valid also in the
presence of ordinary matter with a density $\rho$ and pressure $p$
satisfying the conservation law for the tensor of energy-momentum
\begin{equation*}
    \dot \rho+3H\,(\rho+p)=0,
\end{equation*}
in the isotropic homogeneous expanding universe, so that a finite
density of energy is falling down, and the permitted region of
phantom energy becomes more narrow. The same arguments are valid
for the temporal component of the vector field, too.

In addition, we emphasize that the metric of (\ref{FRW})
substituted into the Einstein--Hilbert action of gravity
\begin{equation*}
    S_{\rm EH} = -\frac{1}{16\pi G}\int R\,\sqrt{-g}\;{\rm
d}^4x
\end{equation*}
after the integration by parts, takes the form
\begin{equation}\label{gravi}
    S_{\rm FRW}= -\frac{3}{8\pi G}\int \dot a^2 a\;{\rm d}^4x,
\end{equation}
which is the action for a scalar field $a(t)$ up to a
normalization factor with a `phantom' sign of kinetic term. That
is why a cosmological singularity is possible. A reason for a
consistency of such the theory of gravity is due to the FRW metric
does not give `free gravitational waves', but it refers to the
coupled case, i.e. we deal with a virtual gravitational field,
while the harmonic oscillations are well behaved. Therefore, the
same features can be ascribed to fields of `gravitational nature',
as we will mean on the vector field.

Next, we consider a vector field, which lagrangian is not
invariant under gauge transformations. For comparison, any gauge
field ${\cal A}_\mu$, for example, an abelian field, contains
unphysical degree, i.e. purely gauge component:
\begin{equation*}
    {\cal A}_\mu=\partial_\mu f.
\end{equation*}
If the energy-momentum tensor of gauge field is gauge invariant,
then the purely gauge component does not propagate. If we add a
gauge-fixing term, the gauge invariance guarantees that the
longitudinal component preserves its bare propagator, i.e. it is
not renormalized. Hence, the longitudinal field undetectable by
gauge interactions remains arbitrary and gauge-dependent, anyway.
So, it would be rather reasonable to study gauge-noninvariant
vector field interacting with gravity, only, in order to touch
main effects due to such the field.

Finally, a current-current interaction
\begin{equation*}
    {\cal L}_{\rm int}=j^\mu{\cal A}_\mu
\end{equation*}
preserves its gauge independence, if the matter current $j^\mu$ is
transversal:
\begin{equation*}
    \nabla_\mu j^\mu \equiv 0.
\end{equation*}
This note could be important, since it gives a chance for
transversal components of the vector field interact with the
matter, if they do not kinematically mixed with longitudinal ones.

Further, the observed quasi-isotropic cosmic microwave background
(CMB) fixes a universe rest frame, i.e it forms an ether. Indeed,
experimentalists measure a velocity of Earth motion in CMB in
order to extract a small true anisotropy (so-called dipole
subtraction). Thus, the expanding Universe determines a specific
rest frame, which allows one to introduce fields, corresponding to
the symmetry of the frame. Particularly, the vector field
preserving the homogeneous and isotropic metric should have a
temporal component, only: $\phi^m=(\phi_0,\boldsymbol 0)$. It was
{\it a priori} clear, that the curved metric of expanding Universe
breaks the Lorentz-invariance, of course. That is why the vector
field components have a fixed form required by the symmetry of the
physical system. In addition, we expect that an effective
potential of the vector field should be of a gravitational origin,
so that its dimensional parameters are posed in the Planck mass
range. If the potential has a stable point, we expect that a
corresponding value of temporal component $\phi_0$ is given by
$\phi_*\sim m_{\rm Pl}$. We will see that the expansion causes a
small time dependence of $\phi_0(t)$ in vicinity of $\phi_*$. So,
the variation of $\phi_0$ at galactic scales of distance and time
is negligible. This fact means that we get a factorization: the
field equations give a true function of $\phi_0(t)$, which can be
considered as a constant external field $\phi_*$ at galactic
level.

Furthermore, given the external vector field source
$\phi_0=\phi_*$, we can allow small stochastic fluctuations of
spatial components $\boldsymbol \phi$, only. These fluctuations
could be related with matter fields. For instance, a complex
scalar matter field $\chi$ can develop a vacuum fluctuations
(condensates) at a characteristic scale $M\ll m_{\rm Pl}$, so that
the spatial derivatives $\boldsymbol \nabla\chi$ coupled to the
spatial components of vector field, can stochastically get nonzero
vacuum expectations
$\langle\partial_i\chi^\dagger\partial_j\chi\rangle$ at the same
scale $M$, and the effective potential of spatial vector field
could acquire a form yielding a stochastically stable point of
$\langle\boldsymbol \phi^2\rangle=\phi_\star^2\sim M^2$, too.
Thus, we arrive at the position with two scales of expectation
values for the vector field: the temporal component $\phi_*$ of
the order of $m_{\rm Pl}$ and the spatial component $\phi_\star$
of the order of $M\ll m_{\rm Pl}$. This small parameter is
characteristic for the problem, and the arrangement of scales is
valid, say, at $M$ determined by a spontaneous symmetry breaking
in a great unification theory, $M\sim M_{\rm GUT}$. The constant
vector field with the scale arrangement serves as the external
source in the gravity equations at the galactic scale.

The consideration is essentially transformed in the spherically
symmetric case, since the symmetry of the system allows the
spatial component directed along the radius-vector: $\boldsymbol
\phi =\phi_\star\,\boldsymbol n$ with the unit vector $\boldsymbol
n=\boldsymbol r/r$. Therefore, the external vector field takes the
form $\phi^m=(\phi_*,\phi_\star\,\boldsymbol n)$, which is purely
gauge field with a gauge function $f(t,r)=\phi_*\,t-\phi_\star\,r$
and $\phi_m =\partial_m f(t,r)$ in a flat space-time limit, so
that this field can be detected gravitationally, only. In
addition, this point gives an extra argument for the lagrangian of
(\ref{v-lagr}), since one could completely substitute a scalar
field derivative for the vector field: $\phi_m =\partial_m
f(t,r)$, so that the potential near the extremal point, say,
$V\approx V_0-\bar \mu^2\, (\phi_0^2-\boldsymbol \phi^2)$, would
be transformed to an ordinary kinetic term of scalar field $f$ up
to a normalization factor, while the other terms represent higher
derivative contributions in an affective lagrangian.

In this paper we study a vector field dynamics in a static
spherically symmetric metric, which should be in halos of spiral
galaxies. Rotational curves in such the galaxies become flat in
the regions of dark halos, that corresponds to $1/r^2$-dependence
of dark matter density on the distance from the galaxy center. We
show that the covariant derivatives of vector fields naturally and
uniquely generate such the dependence. A physical reason for the
conclusion is rather simple. First, the vector field gains a
dynamical mass term determined by a spatial curvature. Second, the
problem introduces a small parameter, a constant velocity of
rotation in the dark halos $v_0$.
Third, following the factorization of cosmological and galactic
scales, at large distances we have to reach a cosmological limit:
the vector field negligibly slowly evolving with time and
distance, i.e. the constant field at the galaxy scale with the
hierarchy of expectations $\kappa=\phi_\star/\phi_*\ll 1$. We find
the relation between two small parameters, the velocity and scale
ratio: $v_0^2=\kappa/\sqrt{2} .$ Then, the curvature should fall
as $1/r^2$, only, that reproduces the flat rotational curves.

The paper is organized as the following: In section 2 we study
Einstein equations with the vector field in a static spherically
symmetric metric
\begin{equation}\label{static}
    {\rm d}s^2 =\mathfrak f(r)\,{\rm d}t^2-\frac{1}
    {\mathfrak h(r)}\,{\rm d}r^2-r^2[{\rm
    d}\theta^2+\sin^2\theta\,{\rm d}\varphi^2],
\end{equation}
by deriving the energy-momentum tensor. Then we remind a condition
following from the observation of flat rotation curves: the
function $\mathfrak f(r)$ should take a specific form up to small
corrections neglected, while $\mathfrak h(r)$ should be close to
1. Then we expand in small $v_0^2$ and $\kappa$ and find the
solution\footnote{We use an ordinary notation for the derivative
with respect to the distance by the prime symbol $\partial_r f(r)
=f^\prime(r)$.}
\begin{equation*}
    v_0^2=\kappa/\sqrt{2},\quad \mathfrak h(r) =
    1-2v_0^2,\quad\mathfrak f'(r)=\frac{2v_0^2}{r},
\end{equation*}
giving a $1/r^2$-profile of the curvature and a flat asymptotic
behavior for a velocity of rotation\footnote{The same metric was
exactly first found by Nucamendi, Salgado and Sudarsky (the last
reference of \cite{expl}). They also studied the light bending and
gravitational lensing, which do not conflict with observations, so
that we will not concern for this question in the present paper.},
so that the velocity squared is determined by the small ratio of
spatial component to the temporal one, i.e. the ratio of
characteristic matter scale to that of gravity, the Planck mass,
that gives $M\sim 10^{-7} m_{\rm Pl}\sim 10^{12}$ GeV, a scale in
the range of GUT breaking\footnote[1]{One can also put $M\sim
\mu^2/m_{\rm Pl}$ yielding $\mu\sim 10^{16}$ GeV, which is closer
to the GUT scale. However, this treatment suffers from `tending to
desirable result', I think.}. In section 3 we analyze the
factorization. Section 4 is devoted to the description of
applicability region for the flat rotation curves in the framework
of constant vector fields, that gives a natural estimate of
Milgrom's acceleration. Other approaches are shortly discussed for
comparison. The results are summarized in Conclusion.

\section{Generic equations}
General expressions for the Christoffel symbols, Ricci tensor and
scalar curvature for the spherically symmetric static metric of
(\ref{static}) are listed in Appendix \ref{metro}.

Then, a tensor entering the Einstein equations
\begin{equation}\label{eins}
    R_{\mu\nu}-\frac{1}{2}\,g_{\mu\nu}\,R=8\pi G\,T_{\mu\nu},
\end{equation}
i.e.,
\begin{equation*}
    G_\mu^\nu =R_\mu^\nu-\frac{1}{2}\,\delta_\mu^\nu\,R,
\end{equation*}
takes the form
\begin{eqnarray}
  G_t^t &=& \frac{1-\mathfrak h}{r^2}-\frac{\mathfrak
  h'}{r},\qquad
  G_r^r = \frac{1-\mathfrak h}{r^2}-\frac{\mathfrak
  f'}{r}\,\frac{\mathfrak f}{\mathfrak h},
  \\[2mm]
  G_\theta^\theta &=&
  G_\varphi^\varphi=-\frac{1}{2}\,\frac{\mathfrak h}{\mathfrak
  f}\,\mathfrak f''-\frac{\mathfrak h}{2 r}\,\left(\frac{\mathfrak f'}{\mathfrak
  f}+
  \frac{\mathfrak h'}{\mathfrak h}\right)+\frac{1}{4}\,\mathfrak
  h\,\frac{\mathfrak f'}{\mathfrak f}\left(\frac{\mathfrak f'}{\mathfrak
  f}-\frac{\mathfrak h'}{\mathfrak h}\right).
\end{eqnarray}

In polar coordinates $x^m=(t,r,\theta,\varphi)$, covariant
derivatives of vector field $\phi^m=(\phi_0,\phi_r,0,0)$ in the
metric of (\ref{static}) are equal to
\begin{equation}\label{cov-stat}
    \begin{array}{ll}
    \displaystyle
      \phi^t_{\;;t} =\frac{1}{2}\,\frac{\mathfrak f'}{\mathfrak f}\,\phi_r, &
      \displaystyle\;\;
      \phi^t_{\;;r} =\phi_0'+\frac{1}{2}\,\frac{\mathfrak f'}{\mathfrak f}\,\phi_0,
      \\[4mm]
    \displaystyle
      \phi^r_{\;;t} =\frac{1}{2}\,{\mathfrak h}\,{\mathfrak f'}\,\phi_0, &
      \displaystyle\;\;
      \phi^r_{\;;r} =\phi_r'-\frac{1}{2}\,\frac{\mathfrak h'}{\mathfrak h}\,\phi_r,
      \\[4mm]
    \displaystyle
      \phi^\theta_{\;;\theta} =\phi^\varphi_{\;;\varphi} =\frac{1}{r}\,\phi_r, &
      \displaystyle
      \sqrt{-g} =\sqrt{{\mathfrak f}/{\mathfrak
      h}}\,r^2\,\sin^2\theta,
    \end{array}
\end{equation}
where we have also shown the determinant of the metric, too.

Squaring $({\cal D}\phi)^2 \equiv
    \phi^m_{\;\;;\mu}\,\phi^n_{\;\;;\nu}\;g_{mn}\,g^{\mu\nu}$ gives
\begin{eqnarray}
    ({\cal D}\phi)^2 &=&
    -\mathfrak f\,\mathfrak h\left(\phi_0'+\frac{1}{2}\,\frac{\mathfrak
    f'}{\mathfrak f}\,\phi_0\right)^2-\frac{1}{4}\,\frac{\mathfrak
    h}{\mathfrak f}\,(\mathfrak f')^2\,\phi_0^2 \nonumber\\[3mm]
\label{D2}
    &&+\frac{1}{4}\left(\frac{\mathfrak f'}{\mathfrak f}\right)^2\,\phi_r^2+
    \left(\phi_r'-\frac{1}{2}\,\frac{\mathfrak h'}{\mathfrak
    h}\,\phi_r\right)^2+\frac{2}{r^2}\,\phi_r^2.
\end{eqnarray}

The action of the vector field in the problem is written down as
\begin{equation*}
    S_{\rm V} =-\frac{1}{2}\int{\rm d}t\, {\rm d}r\,{\rm d}\theta\,{\rm
    d}\varphi\,({\cal D}\phi)^2\,\sqrt{-g},
\end{equation*}
where we have not displayed a potential, which properties have
been described in the Introduction, so that it makes dominant
contribution to the field equations for the vector field, that
result in the mentioned quasi-constant fields as will be discussed
in section \ref{sec3}.

To the moment we remind the restriction following from the
flatness of rotation curves. In this way, a particle motion in the
metric of (\ref{static}) is determined by the Hamilton--Jacobi
equations
\begin{equation}\label{1}
    g^{\mu\nu}\;{\partial_{\mu} S}\,{\partial_{\nu} S} - m^{2} =
    0,
\end{equation}
where $m$ denotes the particle mass. Following the general
framework, we write down the solution in the form, which
incorporates two integrals of the motion in the spherically
symmetric static gravitational field,
\begin{equation}\label{2}
    S = -{\cal E}\, t+{\mathfrak M}\,\theta+{\cal S}(r),
\end{equation}
where $\cal E$ and $\mathfrak M$ are the conserved energy and
rotational momentum, respectively. Then, from (\ref{1}) we deduce
\begin{equation}\label{3}
    \left(\frac{\partial {\cal S}}{\partial r}\right)^2 =
    \frac{1}{\mathfrak f\mathfrak h}\,{\cal E}^2-\frac{1}{\mathfrak h}
    \left(\frac{\mathfrak
    M^2}{r^2}+m^2\right),
\end{equation}
which results in
\begin{equation}\label{4}
    {\cal S}(r) = \int \limits_{r_0}^{r(t)} \textrm{d}r\;
    \frac{1}{\sqrt{\mathfrak f\mathfrak h}}
    \sqrt{{\cal E}^2-V^2(r)},
\end{equation}
where $V^2$ is an analogue of potential,
\begin{equation*}
V^2(r) = \mathfrak f\left(\frac{\mathfrak M^2}{r^2}+m^2\right).
\end{equation*}
The trajectory is implicitly determined by the
equations
\begin{eqnarray}
  \frac{\partial S}{\partial {\cal E}} &=& \textsf{const} = -t
  +\int\limits_{r_0}^{r(t)} \textrm{d}r\; \frac{1}{\sqrt{\mathfrak f\mathfrak h}}
    \frac{\cal E}{\sqrt{{\cal E}^2-V^2(r)}}, \label{p1}\\
  \frac{\partial S}{\partial {\mathfrak M}} &=& \textsf{const} =
  \theta
  -\int\limits_{r_0}^{r(t)} \textrm{d}r\;\frac{1}{\sqrt{\mathfrak f\mathfrak h}}\,
  \frac{\mathfrak f}{r^2}
    \frac{\mathfrak M}{\sqrt{{\cal E}^2-V^2(r)}}.\label{p2}
\end{eqnarray}
Taking the derivative of (\ref{p1}) and (\ref{p2}) with respect to
the time\footnote{As usual $\partial_t f(t) =\dot f$.}, we get
\begin{eqnarray}
  1 &=& \dot r\; \frac{\cal E}{\sqrt{\mathfrak f\mathfrak h}\,
  \sqrt{{\cal E}^2-V^2(r)}}, \\
  \dot \theta &=& \frac{\dot r}{r^2}\;\frac{\mathfrak f}{\sqrt{\mathfrak f\mathfrak
  h}}\,\frac{\mathfrak M}{  \sqrt{{\cal
  E}^2-V^2(r)}},
\end{eqnarray}
and, hence,
\begin{equation}\label{8}
    {\cal E} = \frac{\mathfrak f}{r v}\,{\mathfrak M},
\end{equation}
relating the energy and the rotational momentum, where we have
introduced the velocity
\begin{equation*}
v \stackrel{\mbox{\tiny def}}{=} r\dot \theta.
\end{equation*}
The points of return are determined by
\begin{equation*}
\dot r =0, \quad \Rightarrow\quad {\cal E}^2 - V^2 =
0,\quad\Rightarrow\quad {\mathfrak M}^2 = m^2 r^2
\frac{v^2}{\mathfrak f-v^2}.
\end{equation*}
The circular rotation
takes place, if two return points coincide with each other, i.e.
we have the stability of zero $\dot r$ condition. Introducing a
`proper distance' $\lambda$ by
\begin{equation*}
\frac{\partial}{\partial\lambda} = \frac{\partial
r}{\partial\lambda}\,\frac{\partial}{\partial r} = \sqrt{\mathfrak
f\mathfrak h}\,\frac{\partial}{\partial r}
\end{equation*}
we deduce the wave-equation with spectral parameter ${\cal E}^2$
and `potential' $V^2$
\begin{equation}\label{10}
    \left(\frac{\partial \cal S}{\partial\lambda}\right)^2 = {\cal
    E}^2- V^2,
\end{equation}
so that the stability of circular motion implies the stability of
potential,
\begin{equation}\label{11}
    \frac{\partial V^2}{\partial r} = 0.
\end{equation}
Then, we get
\begin{equation}\label{12}
    v^2 = \frac{1}{2}\, \frac{\textrm{d}\mathfrak f(r)}{\textrm{d}\ln
    r}.
\end{equation}
Introducing a re-scaled velocity with respect to the proper time,
\begin{equation*}
{\mathfrak v}^2 = \frac{1}{\mathfrak f}\, v^2,
\end{equation*}
we get the result of \cite{b}
\begin{equation*}
    {\mathfrak v}^2 = \frac{1}{2}\, \frac{\textrm{d}\ln \mathfrak f}{\textrm{d}\ln
    r}.
\end{equation*}
An accuracy of observations do not allow us to distinguish $v$
from $\mathfrak v$, since $\mathfrak f(r)\to 1$ at large distances
in dark halos. So, in the halo we put
\begin{equation*}
    v =v_0 ={\rm const.}
\end{equation*}
Therefore,
\begin{equation}\label{flim}
    \mathfrak f'= \frac{2v_0^2}{r}
\end{equation}
gives the profile of flat rotation curves\footnote[2]{More
specified consideration can be found in \cite{galact}, for
instance.}. Nonrelativistically, we get
\begin{equation}\label{hlim}
    \mathfrak h =1-q,\quad q\ll 1,
\end{equation}
while the form of dependence on the distance is not fixed by the
flatness.

Let us show that for the vector field serving as an external
source, there is a solution of Einstein equations with
\begin{equation}\label{qlim}
    q'\approx 0,
\end{equation}
where the approximation means the leading order in small parameter
of $v_0^2$, which is, in practice, $v_0^2\sim 10^{-7}$, since the
characteristic velocity in the halos is about $100-150$ km/s, i.e
$v_0\sim (1/3-1/2)\times 10^{-3}$.

Remember, that we consider the following limit for the external
vector field in a spiral galaxy:
\begin{equation}\label{sur}
    \phi_0=\phi_*,\quad \phi_r=\phi_\star,\quad
    \kappa=\frac{\phi_\star}{\phi_*}\ll 1,
\end{equation}
so that
\begin{equation}\label{sur2}
    \phi_0'=0,\quad\phi_r'=0,
\end{equation}
and we refer the system of (\ref{hlim})--(\ref{sur2}) as a
C-surface condition. Then
\begin{equation}\label{D2C}
    ({\cal D}\phi)^2\raisebox{-1mm}{$\big|_{\,\rm C}$}=
    -\frac{1}{2}\,\frac{\mathfrak
    h}{\mathfrak f}\,(\mathfrak f')^2\,\phi_*^2 +
    \frac{1}{4}\left(\frac{\mathfrak f'}{\mathfrak f}\right)^2\,\phi_\star^2+
    \frac{2}{r^2}\,\phi_\star^2,
\end{equation}
so that we get terms quadratic in the field, that implies the
appearance of `induced' mass. This fact is characteristic for the
vector field dynamics in the curved space-time. Combining the
C-surface with (\ref{flim}) gives a $\bar{\rm C}$-surface.

It is a straightforward task to calculate the energy-momentum
tensor for the vector field
\begin{equation*}
    T_\mu^\nu =2\frac{g^{\nu\alpha}}{\sqrt{-g}}\,\frac{\delta
    S_{\rm V}}{\delta g^{\mu\alpha}},
\end{equation*}
so that
\begin{equation*}
    T_t^t =-2\mathfrak f\,\frac{\sqrt{\mathfrak h/\mathfrak
    f}}{r^2\sin\theta}\,\frac{\delta S_{\rm V}}{\delta\mathfrak
    f},\qquad
    T_r^r =2\frac{\mathfrak h}{\mathfrak f}\,\frac{\sqrt{\mathfrak h\mathfrak
    f}}{r^2\sin\theta}\,\frac{\delta S_{\rm V}}{\delta\mathfrak
    h},
\end{equation*}
while the angle components are given by
\begin{equation*}
    T_\theta^\theta = T_\varphi^\varphi = -\frac{\sqrt{\mathfrak h/\mathfrak
    f}}{r^2\sin\theta}\,\frac{\delta S_{\rm V}}{\delta \lambda},
\end{equation*}
where $\delta \lambda$ is the dilatation of $\varphi$:
\begin{equation*}
    \delta_\lambda\varphi =\delta\lambda\cdot\varphi.
\end{equation*}
Direct calculations result in
\begin{eqnarray}
  T_t^t\raisebox{-1mm}{$\big|_{\,\rm C}$} &=& -\frac{1}{2}\,\phi_*^2
   \left[-2\mathfrak h\, \mathfrak f''+\frac{1}{2}\,(\mathfrak f')^2\,
   \frac{\mathfrak h}{\mathfrak f}-\frac{4}{r}\,\mathfrak h\,\mathfrak f'\right]
   \nonumber\\[2mm]
   && -\frac{1}{8}\,\phi_\star^2\left[-\frac{8}{r^2}-
   3\left(\frac{\mathfrak f'}{\mathfrak f}\right)^2+\frac{8}{r}\,
   \frac{\mathfrak f'}{\mathfrak f}+4\,\frac{\mathfrak f''}{\mathfrak
   f}\right].
\end{eqnarray}
Applying $\bar{\rm C}$-surface in the leading order over $v_0\ll
1$ and $\kappa\ll 1$ we get
\begin{equation}\label{00}
    T_t^t\raisebox{-1mm}{$\big|_{\,\bar{\rm C}}$}\approx
    \frac{\phi_*^2}{r^2}\,(2v_0^2+\kappa^2),
\end{equation}
and
\begin{equation}\label{G00}
    G_t^t\raisebox{-1mm}{$\big|_{\,\bar{\rm C}}$}\approx
    \frac{q}{r^2}.
\end{equation}
The Einstein equation for the temporal components reads off
\begin{equation}\label{00=00}
    G_t^t\raisebox{-1mm}{$\big|_{\,\bar{\rm C}}$}=
    8\pi G\,T_t^t\raisebox{-1mm}{$\big|_{\,\bar{\rm
    C}}$}\quad\Rightarrow\quad q=8\pi G\,\phi_*^2\,(2v_0^2+\kappa^2).
\end{equation}
The radial component of energy-momentum tensor equals
\begin{equation*}
    T_r^r\raisebox{-1mm}{$\big|_{\,{\rm
    C}}$}=\frac{1}{4}\,\phi_*^2\,\frac{\mathfrak h}{\mathfrak f}\,
    (\mathfrak f')^2+\frac{1}{r^2}\phi_\star^2,
\end{equation*}
transformed to
\begin{equation}\label{rr}
    T_r^r\raisebox{-1mm}{$\big|_{\,\bar{\rm C}}$}\approx
    \frac{\phi_*^2}{r^2}\,(v_0^4+ \kappa^2),
\end{equation}
while
\begin{equation}\label{Grr}
    G_r^r\raisebox{-1mm}{$\big|_{\,\bar{\rm C}}$}
    =\frac{q-2v_0^2}{r^2},
\end{equation}
and
\begin{equation}\label{rr=rr}
    G_r^r\raisebox{-1mm}{$\big|_{\,\bar{\rm C}}$}=
    8\pi G\,T_r^r\raisebox{-1mm}{$\big|_{\,\bar{\rm
    C}}$}\quad\Rightarrow\quad q-2v_0^2=8\pi
    G\,\phi_*^2\,(v_0^4+\kappa^2).
\end{equation}
Finally, the angle component is equal to
\begin{equation*}
    T_\varphi^\varphi\raisebox{-1mm}{$\big|_{\,{\rm C}}$}=
    T_\theta^\theta\raisebox{-1mm}{$\big|_{\,{\rm C}}$}=
    -\frac{1}{4}\,\frac{\mathfrak h}{\mathfrak f}\,(\mathfrak
    f')^2\,\phi_*^2+\frac{1}{r^2}\,\phi_\star^2,
\end{equation*}
yielding
\begin{equation}\label{tt}
    T_\varphi^\varphi\raisebox{-1mm}{$\big|_{\,\bar{\rm
    C}}$}\approx
    \frac{\phi_*^2}{r^2}\,(\kappa^2-v_0^4),
\end{equation}
so that
\begin{equation}\label{Gtt}
    G_\varphi^\varphi\raisebox{-1mm}{$\big|_{\,\bar{\rm C}}$}=
    \frac{v_0^4}{r^2},
\end{equation}
and
\begin{equation}\label{tt=tt}
    G_\varphi^\varphi\raisebox{-1mm}{$\big|_{\,\bar{\rm C}}$}=8\pi G\,
    T_\varphi^\varphi\raisebox{-1mm}{$\big|_{\,\bar{\rm
    C}}$},\quad\Rightarrow\quad
    v_0^4 =8\pi G\,\phi_*^2(\kappa^2-v_0^4).
\end{equation}
Equations (\ref{00=00}), (\ref{rr=rr}) and (\ref{tt=tt}) are
satisfied at
\begin{eqnarray}
\label{solve-k}
    \kappa =\sqrt{2}v_0^2+{\cal O}(v_0^4),\\[2mm]
    \label{solve-q}
    q ={2}v_0^2+{\cal O}(v_0^4),\\[2mm]
    \label{solve-phi}
    8\pi G\,\phi_*^2 =1+{\cal O}(v_0^2).
\end{eqnarray}
Therefore, this solution results in the temporal component of the
energy-momentum tensor dominates and has the required profile with
the distance:
\begin{equation*}
 T_r^r\sim T_\varphi^\varphi\sim T_\theta^\theta\sim {\cal
 O}(v_0^2)\cdot T_t^t\sim {\cal O}\left(\frac{1}{r^2}\right).
\end{equation*}
Numerically, we get
\begin{equation*}
    \kappa\sim 10^{-7}\quad \Rightarrow\quad M\sim 10^{12}\,\mbox{GeV,}
\end{equation*}
hence, the characteristic scale of matter influence on the vector
field is in the range of GUT and effective scale responsible for
the small neutrino masses.

Thus, the small ratio of two natural energetic scales determines
the rotation velocity in dark galactic halos. In addition, we have
found condition (\ref{solve-phi}), which is analogous to a
definition. Let us test this problem in the next section.

\section{Factorization}\label{sec3}
The previous section treats the vector field as an external
source. That would be valid, if the vector-field equations are
satisfied. In that case we should take into account the filed
potential as we suggest that the characteristic scales in this
potential much greater than the scales induced by the size of
galactic halos as well as the Hubble rate or large-scale
inhomogeneous structures.

So, first, suppose that we deal with the constant isotropic
homogeneous vector field. It means that the radial component is
equal to zero, while the temporal component is posed at the
extremal point of its potential:
\begin{equation*}
    \frac{\partial V}{\partial \phi_0}\raisebox{-1mm}{$\big|_{\,{\rm
    C}}$}=0,\qquad \frac{\partial^2 V}{\partial\phi_0^2}
    \raisebox{-1mm}{$\big|_{\,{\rm C}}$}=m_0^2,\quad
    |m_0|\sim\phi_0=\phi_*\sim m_{\rm Pl}.
\end{equation*}
If we consider an expanding isotropic homogeneous Universe, then
the potential gets an additional term, determined by the Hubble
rate \cite{vecquin}:
\begin{equation*}
    \delta V=\frac{3}{2}\,H^2 \,\phi_0^2,
\end{equation*}
so that the temporal component acquires a slow variation with the
time due to the displacement of stable point, since the field
equation takes the form
\begin{equation*}
    \ddot \phi_0+3H\,\dot\phi_0-3H^2\,\phi_0-\frac{\partial
    V}{\partial \phi_0}=0,
\end{equation*}
and neglecting the time derivative, we get
\begin{equation*}
    3H^2\,\phi_0+\frac{\partial V}{\partial \phi_0}=0.
\end{equation*}
Expanding in $\phi_0$ at $\phi_0=\phi_*+\delta\phi_0$ gives
\begin{equation*}
    3H^2\,\phi_*+(3H^2+m_0^2)\delta\phi_0=0,
\end{equation*}
so that at $H\ll m_{\rm Pl}$
\begin{equation}\label{vary-0}
    \frac{\delta\phi_0}{\phi_*}\approx -\frac{3H^2}{m_0^2},
\end{equation}
which is really small correction, we justify. The induced
time-dependence is due to the Hubble rate
\begin{equation*}
    \delta\dot\phi_0 \approx
    -H\,\phi_*\,6\,\frac{H^2}{m_0^2}\,\left(\frac{1}{H^2}\,
    \frac{\ddot a}{a}-1\right)\quad\Rightarrow\quad |\delta\dot\phi_0/\phi_*
    H|\ll 1.
\end{equation*}

Analogous arguments are valid for the spatial component of the
vector field: the corresponding effective mass of radial component
is $m_r\sim M$, introduced above, while the correction due to the
covariant derivative
\begin{equation*}
    \delta_r V=\frac{2}{r^2}\,\phi_r^2
\end{equation*}
generates a small correction at macroscopic scales $1/r\ll M$, so,
an induced dependence of $\phi_r$ on the distance is actually
suppressed, and $\phi_r\approx\phi_\star$.

If both the expansion and the radial component are nonzero, then
the covariant derivatives
\begin{equation}
\label{if}
    \phi^r_{\;;r}=H\,\phi_0,\qquad
    \phi^\theta_{\;;\theta}=\phi^\varphi_{\;;\varphi}=\frac{1}{r}\,\phi_r+H\,\phi_0,
\end{equation}
induce the correction to the potential
\begin{equation}
\label{if-V}
    \delta_H V
    =\frac{1}{2}\,H^2\,\phi_0^2+\left(\frac{1}{r}\,\phi_r+H\,\phi_0\right)^2,
\end{equation}
which gives similar restrictions on the suppressed variations of
vector field, of course.

Therefore, both the evolution equations and Einstein equations at
galactic scales can be considered with the constant vector field
serving as the external source of gravity.

Note, that as was shown in \cite{vecquin}, the dependence of
$\phi_0$ on time generates a variation of effective gravitational
constant with the time. However, such the dependence should be
small as forced by the experimental observations. We have seen
that in the scheme described above the time-dependence of vector
field is negligible.

Finally, let us discuss the relation
\begin{equation}\label{37}
    8\pi G\,\phi_*^2 =1,
\end{equation}
which looks like a definition, but the constraint. Introduce a
bare gravitational constant $G_0$ and an extra interaction of
vector field with the gravity, so that\footnote{We consider the
FRW metric, while generically instead of $\phi_0^2$ one should
introduce the invariant square of vector field in the interaction,
of course.}
\begin{equation*}
    \frac{1}{16\pi G}\,R\quad\rightarrow\quad \frac{1}{16\pi
    G_0}\,R+\frac{c}{4}\,\phi_0^2\,R.
\end{equation*}
Then, according to \cite{vecquin}, the evolution equation with a
constant vector field $\phi_0=\phi_*$ looks like
\begin{equation*}
    3\left(\frac{1}{8\pi G_0}+\frac{c}{2}\,\phi_*^2\right)H^2
    =\rho-\frac{3}{2}\,H^2\phi_*^2,
\end{equation*}
which reproduces the ordinary form, if we put
\begin{equation*}
    \frac{1}{G}=\frac{1}{G_0}+(1+c)\,4\pi\,\phi_*^2.
\end{equation*}
Taking into account (\ref{37}), we get
\begin{equation*}
    G_0=\frac{2}{1-c}\,G.
\end{equation*}
For instance, at $c=1$ we have $1/G_0=0$, and we find that
$\phi_*^2$ defines the gravitational constant, or the Planck mass.

Thus, we have shown that (\ref{37}) preferably gives a defining
relation for the gravitational constant, which has been used as
the starting point for the scale arrange\-ment. This statement
seems to be justified. On the other hand, it points to a deep
connection of the vector field dynamics with the gravity.

\section{Indicating the Milgrom's acceleration}
As we have just shown in the previous section, the expansion of
Universe causes the time-dependence of metric, which produces a
negligible time-dependence of temporal component of the vector
field. Therefore, the `cosmological limit' of vector field is
consistently reached. However, the radial component of vector
field can induce the angle components of covariant derivatives
(\ref{if}) resulting in the anisotropic potential of (\ref{if-V}).
The anisotropy can be neglected at large distances, when
\begin{equation*}
    \left|\frac{1}{r}\,\phi_r\right|\ll H\,\phi_0.
\end{equation*}
Then, the `cosmological limit' of constant vector field can be
disturbed by the potential of (\ref{if-V}), at distances less than
$r_0$ defined by
\begin{equation}
\label{r0}
    \frac{1}{r_0}\,\phi_r = \varepsilon\,H\,\phi_0\quad \Rightarrow\quad
    \frac{1}{r_0}\,\frac{\phi_\star}{\phi_*} = \varepsilon\,H_0,
\end{equation}
where $H_0=H(t_0)$ is the value of Hubble constant at the current
moment of time $t_0$, and $\varepsilon$ is a parameter of order of
$1-0.1$. Thus, at distances\footnote{The actual parameter is the
ratio of radial component to the distance, of course.} less than
$r_0$ the flatness can be disturbed, since the vector field can
acquire a distance dependence\footnote{The dependence of radial
component should dominate.}. Substituting the ratio
$\phi_\star/\phi_*=\sqrt{2}\,v_0^2$ into (\ref{r0}), we get
\begin{equation}\label{acc}
    \frac{v_0^2}{r_0} =\frac{\varepsilon\,}{\sqrt{2}}\,H_0,
\end{equation}
while the quantity
\begin{equation}\label{a0}
    a_0 =\frac{v_0^2}{r_0}
\end{equation}
is the centripetal acceleration at a `critical radius' of $r_0$.
Then, the critical acceleration is determined by the Hubble
rate\footnote{In units with the speed of light equal to 1.},
\begin{equation}\label{Milgrom}
    a_0 =\frac{\varepsilon\,}{\sqrt{2}}\,H_0,
\end{equation}
and it determines the acceleration below which the limit of flat
rotation curves becomes justified\footnote{In practice,
$\varepsilon\approx 1/4$.}. That is exactly a direct analogue of
the critical acceleration introduced by M.Milgrom in the framework
of modified Newtonian dynamics (MOND) \cite{Milgrom}. In MOND, the
Milgrom's acceleration $a_0$ separates two regimes, the Newtonian
and modified ones, so that at gravitational accelerations $a<
a_0$, the dynamics reaches the limit reproducing the non-Newtonian
flat rotation curves. By empirical data, Milgrom surprisingly
found that $a_0\sim H_0$, which is quite an amazing relation.

Then, equations (\ref{r0})--(\ref{Milgrom}) shows that the scale
of Milgrom's acceleration naturally appears in the framework of
vector field embedded into the theory of gravitation.

Further, we could suppose that in the case, when the gravitational
acceleration produced by the visible matter in the galactic
centers exceeds the critical value, we cannot reach the limit of
flat rotation curves. Indeed, in that case the distance dependence
cannot be excluded for the vector field. The Newtonian
acceleration at the `critical scale' $r_0$ is equal to
\begin{equation*}
    a_0^* =\frac{G {\cal M}}{r_0^2},
\end{equation*}
where $\cal M$ is a visible galactic mass. According to
(\ref{Milgrom}), the critical acceleration is a universal quantity
slowly depending on the time, while (\ref{a0}) implies that the
`critical distance' can be adjusted by variation of parameter
$v_0$. Therefore, we should put
\begin{equation}
\label{coin}
    a_0^* =a_0,
\end{equation}
which yields
\begin{equation}\label{Tully-Fisher}
    v_0^4 = G{\cal M} a_0.
\end{equation}
The galaxy mass is proportional to an H-band luminosity of the
galaxy $L_H$, so that (\ref{Tully-Fisher}) reproduces the
Tully--Fisher law
\begin{equation*}
    L_H \propto v_0^4.
\end{equation*}
Then, other successes of MOND can be easily incorporated in the
framework under consideration, too.

Nevertheless, one could look at (\ref{coin}) as a coincidence,
which can be treated twofold. Indeed, it implies that
\begin{equation*}
    \frac{\phi_\star^2}{\phi_*^2} \sim 
    G{\cal M} H\quad\Rightarrow\quad \phi_\star^2 \sim {\cal M} H.
\end{equation*}
So, first, the extremal point of potential for the radial
component is fixed by the galaxy mass and the Hubble rate, or,
second, if we arrange the scale of $\phi_\star\sim M\sim 10^{12}$
GeV, then the characteristic masses of galaxies should be given by
${\cal M}\sim M^2/H$. The preference for one of two viewpoints is
subjective. Despite of that, a possible challenge is
a coincidence problem for the radial field adjusted to get a
critical acceleration.

For completeness, let us discuss two modern approaches
incorporating the Milgrom's acceleration, too. First, the
empirical suggestion of MOND was reformu\-la\-ted as a
non\-relati\-vistic mechanics in the Lagrange form \cite{aqual},
where the relativistic theory as a scalar-tensor gravity was also
given, though superluminal velocities of the scalar field were
found as well as extragalactic gravitational lensing is too weak.
A `stratified' theory by Sanders \cite{Sanders} involves a priori
constant time-like vector field as a disformal extension of curved
metric to a physical one. Recently, a tensor-vector-scalar theory
for the MOND paradigm succeeded by J.D.Bekenstein \cite{TVS} was
presented as a consistent theory with dynamical fields in
agreement with all observational data. It involves an unknown
function, which guarantees the MOND effect of critical
acceleration in a relativistic theory. In this theory the
Milgrom's acceleration is ad hoc quantity, while in the framework
we have presented its scale is naturally obtained. Note, that our
consideration of Tully--Fisher law has repeated the arguments of
J.D.Bekenstein taken from \cite{TVS}.

The same note concerns for the second approach: a nonsymmetric
gravitation theory (NGT) by J.W.Moffat \cite{NGT}, where the scale
of Milgrom's acceleration has been fixed empirically in order to
extract the model parameters, say, a characteristic scale. The
advantage of NGT is a direct possibility to explicitly fit the
astronomical data on the rotation curves by theoretical formulas,
though the nonsymmetric rank-2 tensor itself (instead of metric)
is rather an exotic object. In addition, the NGT points to an
exponential decrease of rotation velocity at infinitely large
distances.

Thus, in this section we have got a natural estimate of critical
acceleration, the Milgrom's acceleration determining the region of
consistency for the flat rotation curves.

\section{Conclusion}
In this paper we have found that the spherically symmetric static
Einstein equations with the source given by the energy-momentum
tensor of constant vector field have the solution characterized by
the metric corresponding to flat rotation curves in spiral
galaxies at large distances. The feature of such the vector field
is the scale hierarchy: the temporal component is of the order of
Planck mass $m_{\rm Pl}$, while the radial component about $M$
should be suppressed, so that the small parameter $\kappa=M/m_{\rm
Pl}\ll 1$ deter\-mines the square of small rotation velocity
$v_0^2=\kappa/\sqrt{2}$, implying $M\sim 10^{12}$ GeV. The
arrangement of scales should be caused by a specific potential of
vector field, so that its dynamics at cosmological scales is
factorized from the dynamics at galactic scales. Thus, the vector
field can give the explanation for the dark matter in galactic
halos. This statement is enforced by the natural estimate of
Milgrom's acceleration, below which the flatness can be
consistently justified.

We have to point also to a possibility that the vector field,
which spatial components in the asymptotic region is directed
along the radius-vector (in agreement with the symmetry of the
problem), can be a manifestation of a monopole (see papers by
Nucamendi, Salgado and Sudarsky in \cite{expl}). In that case the
magnitude of radial component can fall to zero in vicinity of
galactic centers, that can be exploited for an explanation of
decrease for the dark matter contribution into the rotation
velocity near the galactic centers.

The author is grateful to profs. J.W.Moffat and J.D.Bekenstein for
electronic messages with useful remarks and questions, which allow
him to essentially improve the revised version of this paper.

This work is partially supported by the grant of the president of
Russian Federation for scientific schools NSc-1303.2003.2, and the
Russian Foundation for Basic Research, grant 04-02-17530.

\setcounter{section}{0}
\def\thesection{\Alph{section}}
\section{Appendix: 
Metric values} \label{metro} The spherically symmetric metric of
(\ref{static}) produces the following Christoffel
symbols:
\begin{equation}\label{Christof}
\begin{array}{llllllll}
  \Gamma^t_{tr} = \displaystyle \frac{1}{2}\,\frac{\mathfrak f^\prime}{\mathfrak f},&  &
  \Gamma^r_{rr} = \displaystyle -\frac{1}{2}\,\frac{\mathfrak h^\prime}{\mathfrak h}, &  &
  \Gamma^r_{tt} = \displaystyle \frac{1}{2}\,{\mathfrak f^\prime}\,{\mathfrak h},\\[4mm]
  \Gamma^\theta_{\theta r} = \Gamma^\varphi_{\varphi r} = \displaystyle \frac{1}{r},
  && \Gamma^r_{\theta\theta}=-r\,\mathfrak h, &&
  \Gamma^r_{\varphi\varphi}=-r\sin^2\theta\,\mathfrak h, \\[4mm]
  \Gamma^\theta_{\varphi\varphi}= - \sin\theta\,\cos\theta,&&
  \displaystyle\Gamma^\varphi_{\theta\varphi}
  =\frac{\cos\theta}{\sin\theta},&&
\end{array}
\end{equation}
while the symbols are symmetric over the contra-variant indices,
and other symbols not listed above are equal to zero. In
eqs.(\ref{Christof}) we do not explicitly show the dependence of
metric components on the distance.

Then, the non-zero elements of Ricci tensor are given by
\begin{equation}\label{Ricci}
\begin{array}{lll}
 \displaystyle  R_{tt} = \frac{1}{2}\,\mathfrak h\,\mathfrak
 f''+\frac{1}{r}\,\mathfrak h\,\mathfrak f'-\frac{1}{4}\left(
 \frac{\mathfrak f'}{\mathfrak f}-\frac{\mathfrak h'}{\mathfrak
 h}\right)\,\mathfrak h\,\mathfrak f',
 &  & \\[6mm]
 \displaystyle R_{rr} = -\frac{1}{2}\,\frac{\mathfrak f''}{\mathfrak
 f}-\frac{1}{r}\,\frac{\mathfrak h'}{\mathfrak
 h}+\frac{1}{4}\left(
 \frac{\mathfrak f'}{\mathfrak f}-\frac{\mathfrak h'}{\mathfrak
 h}\right)\,\frac{\mathfrak f'}{\mathfrak f}, &&\\[6mm]
 \displaystyle R_{\theta\theta} = 1-\mathfrak h-r\,\mathfrak
 h'-\frac{1}{2}\,r\,\mathfrak h\left(\frac{\mathfrak f'}{\mathfrak
 f}-\frac{\mathfrak h'}{\mathfrak h}\right),&& R_{\varphi\varphi}
 =R_{\theta\theta}\,\sin^2\theta,
\end{array}
\end{equation}
while the scalar curvature is equal to
\begin{equation}\label{curvature}
    R = \frac{\mathfrak h}{\mathfrak f}\,\mathfrak f''-\frac{1}{2}\,
    \frac{\mathfrak h}{\mathfrak f}\,\left(\frac{\mathfrak f'}{\mathfrak f}-
    \frac{\mathfrak h'}{\mathfrak h}\right)\,\frac{\mathfrak f'}{\mathfrak f}+
    2\frac{\mathfrak h}{r}\left(\frac{\mathfrak f'}{\mathfrak f}+
    \frac{\mathfrak h'}{\mathfrak h}\right)-\frac{2}{r^2}\,(1-\mathfrak
    h).
\end{equation}


\end{document}